\documentclass[twocolumn]{aastex631}

\newcommand{\PS}[0]{\ensuremath{(\boldsymbol{P}_{\alpha} \cdot \nabla)\cdot \mathbf{u}_{\alpha}}}

\newcommand{\ptheta}[1]{{\ensuremath{p_{#1}\theta_{#1}}}}

\usepackage{amsmath}

\usepackage{graphicx} 

\usepackage{url} 
\usepackage{hyperref}
\usepackage{todonotes}
\hypersetup{
     colorlinks = true,
     pdfborder={0 0 0},
     bookmarksopen=true,
     linktocpage=true,
     linkcolor=blue,
     citecolor=blue,
     urlcolor=blue,
}

\begin{document}

\title{Scale Separation Effects on Simulations of Plasma Turbulence}

\author[0009-0003-6942-5410]{Jago Edyvean}
\affiliation{Victoria University of Wellington, Kelburn, 6021, NZ}

\author[0000-0003-0602-8381]{Tulasi N. Parashar}
\affiliation{Victoria University of Wellington, Kelburn, 6021, NZ}

\author{Tom Simpson}
\affiliation{OpenStar Technologies}

\author{James Juno}
\affiliation{Princeton Plasma Physics Laboratory, Princeton, New Jersey 08540, USA}

\author{Gian Luca Delzanno}
\affiliation{Los Alamos National Laboratory, Los Alamos, New Mexico 87545, USA}

\author{Fan Guo}
\affiliation{Los Alamos National Laboratory, Los Alamos, New Mexico 87545, USA}

\author{Oleksandr Koshkarov}
\affiliation{Los Alamos National Laboratory, Los Alamos, New Mexico 87545, USA}

\author{William H Matthaeus}
\affiliation{University of Delaware, Newark, Delaware 19716, USA}

\author{Michael Shay}
\affiliation{University of Delaware, Newark, Delaware 19716, USA}

\author{Yan Yang}
\affiliation{University of Delaware, Newark, Delaware 19716, USA}

\begin{abstract}
Understanding plasma turbulence requires a synthesis of experiments, observations, theory, and simulations. In the case of kinetic plasmas such as the solar wind, the lack of collisions renders the fluid closures such as viscosity meaningless and one needs to resort to higher order fluid models or kinetic models. Typically, the computational expense in such models is managed by simulating artificial values of certain parameters such as the ratio of the Alfv\'en speed to the speed of light ($v_A/c$) or the relative mass ratio of ions and electrons ($m_i/m_e$). Although, typically care is taken to use values as close as possible to realistic values within the computational constraints, these artificial values could potentially introduce unphysical effects. These unphysical effects could be significant at sub-ion scales, where kinetic effects are the most important. In this paper, we use the ten-moment fluid model in the Gkeyll framework to perform controlled numerical experiments, systematically varying the ion-electron mass ratio from a small value down to the realistic proton-electron mass ratio. We show that the unphysical mass ratio has a significant effect on the kinetic range dynamics as well as the heating of both the plasma species. The dissipative process for both ions and electrons become more compressive in nature, although the ions remain nearly incompressible in all cases. The electrons move from being dominated by incompressive viscous like heating/dissipation, to very compressive heating/dissipation dominated by compressions/rarefactions. While the heating change is significant for the electrons, a mass ratio of $m_i/m_e \sim 250$ captures the asymptotic behaviour of electron heating. 

\end{abstract}






\section{Introduction}\label{Ch:Intro}
Many plasma systems in the universe are thought to be collisionless or weakly collisional \citep{quataertAPJ1998, sharma2007electron, MarschLRSP2006}. These plasmas have been observed, or are inferred, to be turbulent. One such natural plasma, that we can study in-situ, is the solar wind, making it an ideal natural laboratory for the purpose of understanding plasma turbulence \citep{MarschLRSP2006, brunocarbone2016book}. The lack of frequent collisions in such systems implies that the traditional fluid closures are no longer applicable and one needs to resort to kinetic approaches \citep{freidberg1982RMP, MarschLRSP2006, kiyani2015dissipation, smith2021driving}. How dissipation occurs in such systems has been the subject of an ongoing debate \cite[e.g.][and references therein]{kiyani2015dissipation}. 

Dissipation in kinetic plasmas has been studied using observations as well as simulations \cite[e.g.][and references therein]{parasharPoP2009,markovskii2010effect, osmanPRL2011, tenbargeAPJL2013, chasapis2015APJ, kiyani2015dissipation, matthaeus2021turbulence}. Fully kinetic simulations, that treat ions and electrons as kinetic Vlasov species, are computationally very expensive and hence typically utilize artificial values of the speed of light($c$) and the relative ion-electron mass ratio ($m_i/m_e$). Artificially low speed of light softens the time-step requirement to resolve light waves and other fast dynamical phenomena. Artificially heavy electrons reduce the separation between the ion and electron kinetic scales, reducing the grid resolution requirements. Some simulations use realistic proton-electron mass ratio \cite[e.g.][]{tenbargeAPJL2013, gary2016ApJ} by compromising the inertial range bandwidth. A realistic mass ratio necessarily limits such simulations to a system size that can not capture the self-consistent inertial range dynamics of plasma turbulence \citep{parasharApJ15}. The ions are almost in the decoupled regime and hence the conclusions about their dynamics and heating from such small simulations are suspect. On the other hand, in simulations that use artificial $m_i/m_e$ \cite[e.g.][]{matthaeusApJ2020} the electrons appear to start heating at ion scales. This, however, was also observed in MMS data recently by  \textcite{manzini2024}. 

\begin{table*}[ht!]
\begin{tabular}{||c |c |c |c |c |c |c|c |c |c |c | c ||}
 \hline
 $\boldsymbol{\omega_{pe}}$ & $\boldsymbol{v_{Ai}/c}$ & $\boldsymbol{du/v_{Ai}}$ & $\boldsymbol{dt\omega_{pe}}$& $\boldsymbol{eps}$ & $\boldsymbol{\beta}$ &  $\boldsymbol{L/d_i}$&$\boldsymbol{d_e}$ & $\boldsymbol{n_0}$ & $\boldsymbol{c}$ & $\boldsymbol{\beta_i/\beta_e}$&$\boldsymbol{N_x,N_y}$\\ 
 \hline
  1 & 0.0115 & 0.2 & 0.2 & 0.2 & 0.1 &  8$\pi$&1 & 1 & 1 & 1   &2048\\ 
 \hline 
\end{tabular}
\caption{Parameters kept constant across all 10-moment two-fluid Gkyell simulations.}
\label{tab:paramconst}
\end{table*}

\begin{table*}[ht!]
\centering
\begin{tabular}{|c |c |c |c | c | c | c | c | c | c | c|c | c | c|}
 \hline
 $\boldsymbol{m_{i} / m_{e}}$ & $\boldsymbol{v_{Ae}}$ &  $\boldsymbol{B_{0}}$  & $\boldsymbol{dB}$  & $\boldsymbol{\omega_{ce}}$ & $\boldsymbol{\omega_{ci}}$ & $\boldsymbol{d_{i}}$  & $\boldsymbol{L}$ & $\boldsymbol{\tau_{0}}$ & $\boldsymbol{dx}$  \\ 
 \hline
 25   & 5.74$\times 10^{-2}$ & 5.74$\times 10^{-2}$  & 1.15$\times 10^{-2}$ & 5.74$\times 10^{-2}$ & 2.30$\times 10^{-3}$  & 5    &  125.66  & 8.7$\times 10^{3}$ &  6.14$\times 10^{-2}$  \\ 
 \hline
 50   & 8.13$\times 10^{-2}$ & 8.13$\times 10^{-2}$ & 1.63$\times 10^{-2}$ & 8.13$\times 10^{-2}$ & 1.63$\times 10^{-3}$ & 7.07 &  177.72  & 1.23$\times 10^{4}$ & 8.68$\times 10^{2}$  \\
 \hline
 100  & 1.12$\times 10^{-1}$ & 1.12$\times 10^{-1}$ & 2.30$\times 10^{-2}$ & 1.12$\times 10^{-1}$ & 1.15$\times 10^{-3}$  & 10    &   251.33   &  1.74$\times 10^{4}$ & 1.23$\times 10^{-1}$  \\
 \hline
 250  & 1.81$\times 10^{-1}$ & 1.81$\times 10^{-1}$ & 3.64$\times 10^{-2}$ & 1.81 $\times 10^{-1}$ & 7.27$\times 10^{-4}$ & 15.8 &  397.38  &  2.75$\times 10^{4}$ & 1.94 $\times 10^{-1}$  \\
 \hline
 500  & 2.57$\times 10^{-1}$ & 2.57$\times 10^{-1}$ & 5.14$\times 10^{-2}$ & 2.57$\times 10^{-1}$ & 5.14$\times 10^{-4}$ & 22.4 &  561.99  &  3.89$\times 10^{4}$ & 2.74$\times 10^{-1}$  \\
 \hline
 1000 & 3.64$\times 10^{-1}$ & 3.64$\times 10^{-1}$ & 7.27$\times 10^{-2}$ & 3.64$\times 10^{-1}$ & 3.64$\times 10^{-4}$ & 31.6 &  794.77  &  5.50$\times 10^{4}$ & 3.88$\times 10^{-1}$  \\
 \hline 
 1836 & 4.93$\times 10^{-1}$ & 4.93$\times 10^{-1}$ & 9.86$\times 10^{-2}$ & 4.93$\times 10^{-1}$ & 2.68$\times 10^{-4}$ & 42.8  &  1076.90  &  7.45$\times 10^{4}$ & 5.26$\times 10^{-1}$  \\ 
 \hline 
\end{tabular}
\caption{Parameters for all 10-moment two-fluid Gkyell simulations. where the electron Alfv\'en velocity is $v_{Ae} [c]$, the background field strength is $B_{0} [c$ $m_{e}^{1/2}\mu_0^{1/2}n_0^{1/2}]$, magnetic field fluctuations is $dB [c$ $m_{e}^{1/2}\mu_0^{1/2}n_0^{1/2}]$, electron cyclotron frequency is $\omega_{ce} [\omega_{pe}]$, ion cyclotron frequency is $\omega_{ci} [\omega_{pe}]$, ion inertial length is $d_i [d_e]$, domain length is $L [d_e]$, eddy turnover time is $\tau_0 [d_e c^{-1}]$ and grid size is $dx [d_e]$.}
\label{tab:param}
\end{table*}

The effect of artificial values of $c$ and $m_i/m_e$ on the turbulent dynamics in such simulations have received little attention. Some work has been done in the context of reconnection, fusion, and linear theory \citep{daughton2006PoP, guo2007CPL, le2013PRL, liu2015JGRSP,howard2015PPaCF, bernard2019PoP, li2019APJ, verscharenRAS2020}. In linear theory, the kinetic Alfv\'en waves (KAWs) get damped heavily well before reaching electron scales for $m_i/m_e > 125$ \citep{verscharenRAS2020}. In the context of reconnection, the electron diffusion region elongates with increasing mass ratio as $(m_i/m_e)^{1/4}$ \citep{daughton2006PoP, le2013PRL}. The orientation of the x-line is also affected by the mass ratio in the case of asymmetric reconnection \citep{liu2015JGRSP}. Although many properties such as the reconnection rate and ion heating do not change significantly with the mass ratio, the dynamics of the electrons are significantly affected by it \citep{li2019APJ}. 

In the context of turbulence, \cite{gary2016ApJ} conducted 3D particle-in-cell simulations of whistler turbulence, varying the mass ratio. They found that the heating of electrons does not change significantly, while the heating of the ions decreases as the mass ratio is increased to a realistic value. The increasing mass ratio in their simulations, however, corresponded to increasing ion scales, with the ion gyro-radius quickly becoming comparable to the system size, pushing the energy containing wavenumber well below $kd_i\sim 1$, where $k$ is wavenumber and $d_i$ is ion inertial length.

A complication introduced by reduced parameters is the relevance of various scales in dissipative dynamics of the kinetic range turbulence. For example, although the heat flux for tokamak discharges shows significant build-up at electron scales for well resolved simulations, the overall value is sensitive to the mass ratio \citep{howard2015PPaCF}. Using gyrokinetic simulations with realistic ion-electron mass ratio, \cite{toldPRL2015} found that the electron collisional dissipation peaks close to $k_\perp\rho_i\sim 1$ and the ion collisional dissipation peaks near $k_\perp\rho_i \sim 20$. Although one can expect electrons to heat at ion scales via Landau damping in gyrokinetics, such scale separation could very well be an artefact of the collisional model used. It also begs the question of whether this process is independent of the physical model. Using fully kinetic simulations \cite{matthaeusApJ2020} also showed that the electron dissipation starts at scales slightly larger than the ion inertial scale. The mass ratio in those simulations however was artificially small ($m_i/m_e=25$) and could partly be responsible for pushing the electron dissipation to ion scales. A systematic study of $m_i/m_e$ variation, keeping as many parameters constant as possible is necessary to start addressing some of these questions. 

In this paper we perform systematic numerical experiments, using two-fluid ten-moment Gkeyll code, varying $m_i/m_e$ from an artificial value of 25 to a realistic value of 1836 for proton-electron plasma. We study the properties of kinetic range turbulence including the spectral features, plasma heating, kurtosis, and compressibility. In section \ref{simulations} we describe the simulation setup. Section \ref{Results} describes the electron mass dependence of various turbulence properties and section \ref{Conculsion} concludes by summarizing our results and discussing planned future work into the scale dependence of dissipation as a function of $m_i/m_e$. In order to support the findings of this paper kinetic tools such as fully kinetic simulations \citep{parasharApJ15, juno2018discontinuous}, or the Spectral Plasma Solver (SPS) \citep{delzanno2015JCP, vencels2016JoPCS, roytershteyn2018FASS,koshkarov2021CPC}, could be used to bridge the gap between kinetic and fluid models. That however, is a computationally challenging endeavour and will be addressed in future studies. 

\section{Simulation Setups}\label{simulations}
The moments of the Vlasov equation give us the following equations \citep{hakimJFE2008}, for a given species (species label suppressed for brevity)

\begin{eqnarray}
     \frac{\partial n}{\partial t} + \frac{\partial}{\partial x_j}(nu_j)& = & 0 \\
     m \frac{\partial}{\partial t}(nu_i) + \frac{\partial \mathcal{P}_{ij}}{\partial x_j} & = & nq(E_j + \epsilon_{ijk} u_j B_k)  \\
    \frac{\partial \mathcal{P}_{ij}}{\partial t} + \frac{\partial \mathcal{Q}_{ijk}}{\partial x_k}& = & nqu[iE_j] + \frac{q}{m} \epsilon_{[ikl}\mathcal{P}_{kj]}B_l
\end{eqnarray}

where $q$ is charge, $m$ is mass, $n$ is number density, $u$ is flow velocity, $E$ and $B$ are the electric and magnetic field respectively, $\mathcal{P}_{ij} \equiv m\int v_i v_j f d^3 v  $ is the second moment of the distribution function, $\mathcal{Q}_{ijk} \equiv m \int v_i v_j v_k f d^3 v $ is the third moment of the distribution function, and the square brackets denote a sum over permutations of the indices. The third moment tensor can be written $\mathcal{Q}_{ijk} = q_{ijk} + u_{[i} \mathcal{P}_{jk]} - 2mnu_iu_ju_k$, where $q_{ijk} \equiv m \int (v_i-u_i)(v_j-u_j)(v_k-u_k)fd^3v$ is the heat flux tensor. The fluid equations are coupled via Maxwell's equations. 

The 10 moment model has seen a lot of successes in previous studies such as the work done by \textcite{ngPoP2017} on island coalescence, \textcite{ng2019JGRSP} on kinetic drift instabilities and \textcite{tenbarge2019JGRSP} on 3D asymmetric reconnection, which is turbulent in the direction perpendicular to the reconnection plane. In the study by \textcite{tenbarge2019JGRSP}, they compared the 10 moment model with kinetic particle-in-cell simulations as well as MMS data and found that the model captures a good amount of kinetic dynamics. These kinetic validations make this model an enticing choice for the study of kinetic range dynamics while reducing the required computational load. Thus, this model is optimally positioned to tackle the problem of ion-electron scale separation effects.

The set of equations is closed by choosing a reasonable model for the heat flux tensor. In the collisionless limit, a three-dimensional extension of the Hammett-Perkins closure is used \citep{ngPoP2017}

\begin{equation}
    q_{ijk}(\mathbf{x}) = n(\mathbf{x})\hat{q}_{ijk}(\mathbf{x}),
\end{equation}

where $\hat{q}_{ijk}$ in Fourier space is $\Tilde{q}_{ijk}$ and calculated by

\begin{equation}
    \Tilde{q}_{ijk} = -i\frac{v_t}{|k|}\chi k_{[i}\Tilde{T}_{jk]}
\end{equation}

where $\Tilde{T}_{jk}$ is the Fourier transform of the deviation of the local temperature tensor from the mean, $k$ is the closure parameter, $v_t$ is the thermal velocity and $\chi=\sqrt{4/9\pi}$ is the best fit value for the diagonal $q_{iii}$ component. Through approximation \cite[see][]{wangPoP15} we end up with a simplified closure model, 

\begin{equation}
    \partial_m q_{ijm} \approx v_t |k_0|(\mathbf{P}_{ij}-p\delta_{ij})
\end{equation}

The closure parameter used in this study is constant $k_0=1/d_e$ for both ions and electrons. A convergence study was performed, varying $k_0$ values to $1/d_i$ and $1/d_e$ for both species, to make sure that the choice of this parameter does not significantly affect the results. These ten-moment equations are coupled with the Maxwell's equations and solved by the dimensional- splitting finite-volume method in the ten-moment Gkeyll code. The code uses normalization of the electron inertial length $d_e$ for lengths, the speed of light $c$ for velocities, and inverse electron plasma frequency $\omega_{pe}^{-1}$ for times. The simulations are ran on a domain with periodic boundary conditions.


In order to perform controlled numerical experiments that retain the same inertial range dynamics, while keeping the numerical parameters also the same, we choose to fix the system size in terms of ion inertial length ($d_i=c/\omega_{pi}$), with exactly the same number of grid points across all simulations, giving us the same grid spacing in units of $d_i$. Therefore, the number of grid points is dictated by the realistic mass ratio case and the electron scales are larger and significantly over-resolved for the smallest mass ratio case. This, with each simulation having a domain length of $L=8\pi d_i$ and 2048$^2$ grid points, allows the same amount of the inertial range to be captured.  To enable comparisons with other works, we fix the ratio of the ion Alfv\'en speed to the speed of light. Changing mass ratio means that the electron Alfv\'en speed changes, becoming roughly half the speed of light in the realistic case. Ideally, this requires a relativistic treatment, but that is beyond the scope of the present study.

\begin{figure*}
    \centering
    \includegraphics[width=7.0in]{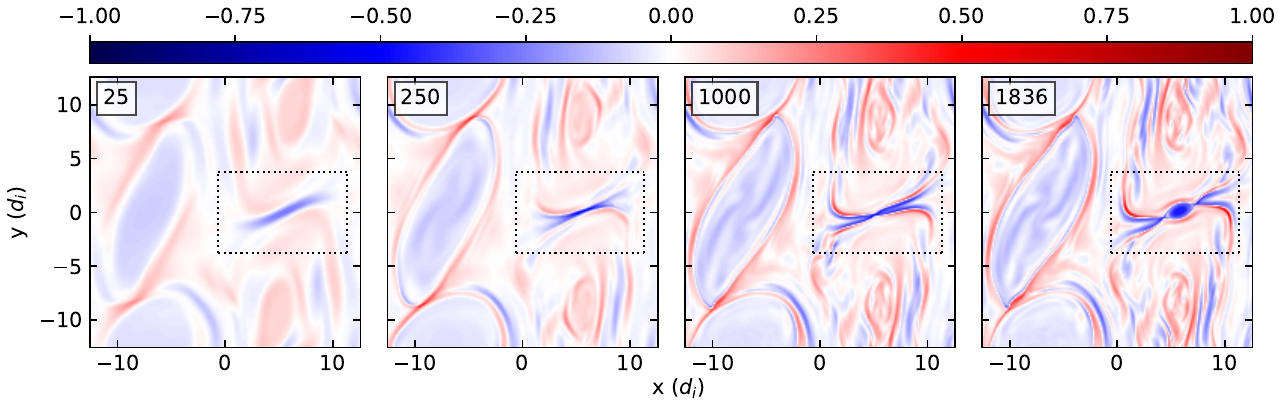}
    \caption{Showing the out of plane current density at $\sim$4.9$\tau_0$ for simulations where $m_i/m_e =$25, 250, 1000, 1836. The periodic domain has been rolled by $4\pi d_i$ in the y direction and $1.76\pi d_i$ in the x direction to place the most intense current sheet near the middle for better visibility. The values have been normalized by 1.5 times the maximum current density value for the snapshots shown. The dotted box region highlights an interesting current sheet shown in fig \ref{fig:jz_zoomed}.}
    \label{fig:jz}
\end{figure*}

The simulations are performed in the 2.5D geometry that keeps three components for all vector quantities, e.g. $\boldsymbol{u}=(u_x, u_y, u_z)$, but retains only two spatial, $\boldsymbol{r}=(x, y)$, dimensions. The parameters kept constant across all simulations are given in the table \ref{tab:paramconst}, where $\omega_{pe}$ is the electron plasma frequency, $v_{Ai}$ is the ion Alfv\'en velocity, $du$ is the velocity fluctuation strength, $dt$ is the time increment, $eps$ is the turbulent fluctuation strength, $\beta$ is the plasma beta, $d_e$ is the electron inertial length, $n_0$ is the number density, $c$ is the speed of light, $\mu_0$ is the permeability of free space, $\epsilon_0$ is the permittivity of free space and $\beta_i/\beta_e$ is the ratio of ion and electron plasma betas. The simulations use the Orszag-Tang vortex as the initial condition, \citep{orszagJFM79}, to allow maximal use of the available bandwidth in the limited inertial range of these simulations \citep{parasharPoP2009,parasharApJ15,li2019JPP}. The simulations run for $6\tau_0$ and are in a fully developed turbulence regime beyond roughly $4\tau_0$, where $\tau_0=L/(2\pi du)$ is the eddy turnover time.


\begin{figure}[h!]
    \centering
    \includegraphics[width=0.8\columnwidth]{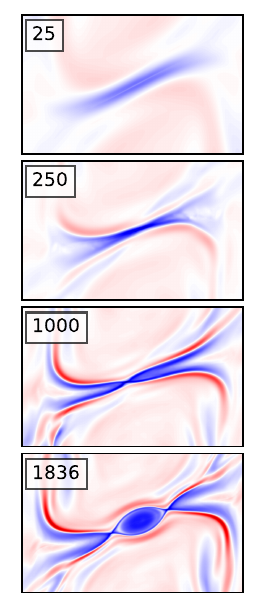}
    \caption{Showing the region of $j_z$ enclosed by the dotted boxes in fig \ref{fig:jz}. The figure acts to qualitatively show the development of structure with increasing $m_i/m_e$, with the colours corresponding to the values shown on the colour bar in fig \ref{fig:jz}.}
    \label{fig:jz_zoomed}
\end{figure}

\section{Results}\label{Results}

We begin by comparing the out of plane current density, $J_z$ for various mass ratios. Fig \ref{fig:jz} shows $J_z$ for mass ratios 25, 250, 1000, and 1836 at a late time ($t\approx 4.9\tau_0$). The periodic domain has been rolled (translated) by $4\pi d_i$ in the y direction and $1.76\pi d_i$ in the x direction to place the most intense current sheet close to the middle of the domain. The values have been normalized by 1.5 times the maximum current density values for the snapshots shown. As expected from decreasing electron scales with increasing mass ratio, the current sheets get thinner. The large scale statistical behaviour is roughly the same for larger mass ratios but the details of individual current sheets vary significantly. 

\begin{center}
\begin{figure}
    \includegraphics[width=3.5in]{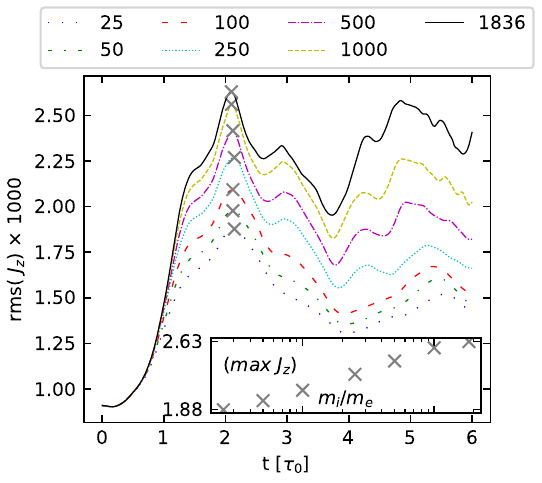}
    \caption{Showing the root-mean-squared, rms, out of plane current density across all simulations.}
    \label{fig:rms_jz}
\end{figure}
\end{center}

Fig \ref{fig:jz_zoomed} shows a zoom-in on the most intense current sheet in the simulation box. The thinning of the current sheet is very evident. For smaller mass ratio, the current sheet is very diffused and Sweet-Parker-like \citep{parkerJGR57, sweetNC58, parkerApJ63}, although faint hints of a bifurcation can be seen. The thinning current sheet with increasing mass ratio also starts to show signs of Petschek-like magnetic reconnections \citep{petschek1964}. For the realistic mass ratio case it becomes unstable to multiple reconnection sites and develops a plasmoid in the middle \citep{bhattacharjeePoP09}. One of the key advantages of the 10 moment model over traditional fluid models is the accuracy of reconnection dynamics, due to the retention of the complete pressure tensor. This has been tested with comparisons to kinetic particle-in-cell models \citep{wangPoP15}, with the 10 moment model capturing a large range of kinetic dynamics.

The increasing intensity of the current sheets with mass ratio can be quantified by studying the time evolution of the root-mean-squared (rms) value of $J_z$. fig \ref{fig:rms_jz} shows the rms out of plane current density as a function of time for various mass ratios. Two important features stand out. First, the peak value of the rms current increases with increasing mass ratio. This is to be expected as smaller mass ratio means smaller electron scales and hence sharper current-sheets that can collapse to electron scales. The insert shows the peak value of rms($J_z$) as a function of mass ratio. rms($J_z$) for the smallest mass ratio is roughly 72\% of the realistic mass ratio. A clear logarithmic increase in the peak current density is also observed. Second, the late time behaviour of $J_z$ varies drastically with mass ratio. Smaller mass ratios show a minor bump in $J_z$ at late times ($\sim 5 \tau_0$) while the larger mass ratio builds up currents as large as at early times. This happens because of a late time reconnection event. As we show later on, the electrons heat less with increasing mass ratio and the simulations with larger mass ratio can sustain a more energetic reconnection event. It is also worth noting that there is a delay in the second bump as mass ratio decreases. This is perhaps expected due to the dynamics of thinning current sheets. As the mass ratio is decreasing, the electrons are becoming heavier. This means the electrons will have more inertia and so move slower, leading to a delay in the bump seen at late times.  

\begin{center}
\begin{figure}
    \includegraphics[width=3.5in]{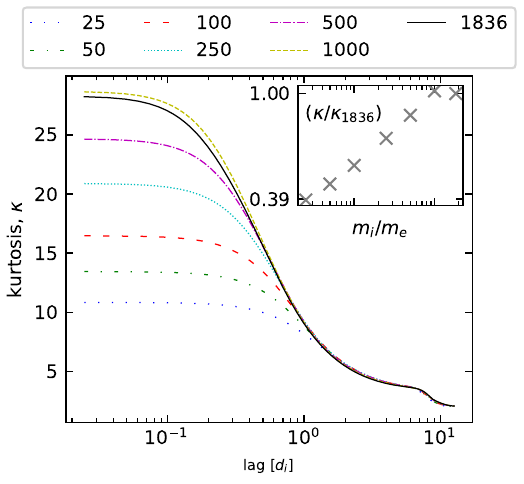}
    \caption{Scale Dependent Kurtosis of $b_x$ as a function of lag in the x direction for various mass ratios. The inset shows the value of kurtosis at a lag of 2$dx$.}
    \label{fig:sdk}
\end{figure}
\end{center}

The localization of the intense current sheets can be quantified using scale dependent kurtosis for the magnetic field. Fig \ref{fig:sdk} shows the scale dependent kurtosis for $b_x$ as a function of lag in the x direction at $4.9\tau_0$. The kurtosis at large scales has a sub-Gaussian value of $\sim 2$ and increases with decreasing lag for all mass ratios. The kurtosis likely starts at $\sim 2$ because of large scale in-homogeneity due to the Orszag-Tang vortex. The kurtosis becomes $\sim 3$ at roughly $7d_i$ signalling the start of the homogeneous turbulence regime. The kurtosis starts plateauing at smaller and smaller scales. The saturated value of the kurtosis increases from 10.8 to 28.6 with increasing mass ratio. The insert in the fig \ref{fig:sdk} shows the kurtosis for different mass ratios, for a lag of 2$dx$. The $m_i/m_e = 1000$ case has a slightly higher kurtosis than the realistic mass ratio, likely because the realistic mass ratio develops plasmoids in the unstable current sheets. The numbers however are fairly close and can be considered to be saturated. Hence, from the perspective of capturing the thin current-sheet dynamics, a mass ratio of greater than 1000 appears desirable.

\begin{figure}[h!]
    \centering
    \includegraphics[width=\columnwidth]{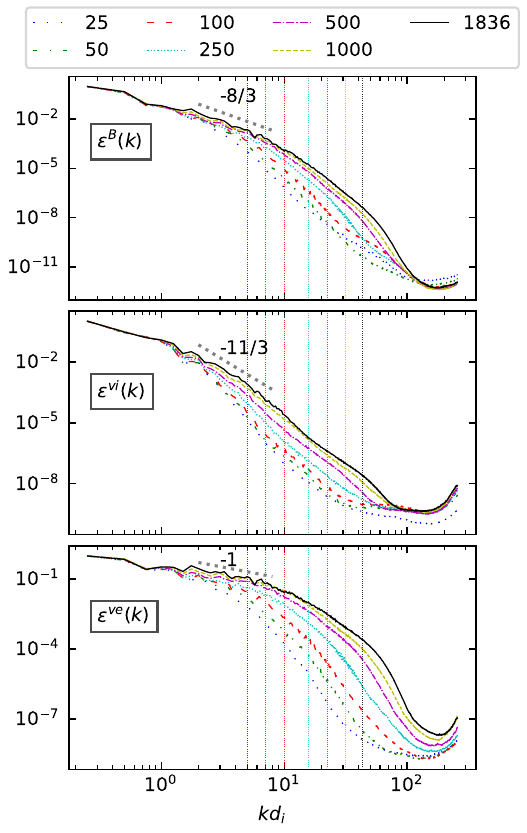}
    \caption{The omni-directional spectra, computed at $t=4.9\tau_0$ for the magnetic field (top), the ion velocity (middle), and the ion velocity (bottom). The coloured dashed lines show the electron inertial length for the different simulations, with the colours represented in the legend. Representative slope lines are shown in each panel. A clear transition to enhanced power at small scales is seen with increasing mass ratio.}
    \label{fig:energy_spectra}
\end{figure}

The intense small-scale structures are fed by a cascade that pumps energy into the small scales. In fig \ref{fig:energy_spectra} we show the power spectra for the magnetic field, ion flow velocity, and electron flow velocity. The vertical dotted lines show $kd_e=1$, identifying the start of sub-electron scale range. The spectra for all the variables overlap at scales larger than the ion scale ($kd_i < 1.$). At smaller scales ($kd_i > 1.$), the spectra start diverging, with the larger mass ratio simulations showing increased power at smaller scales. The electron and magnetic field spectra show exponential roll-over at the smallest scales.

\begin{figure}[h!]
    \centering
    \includegraphics[width=\columnwidth]{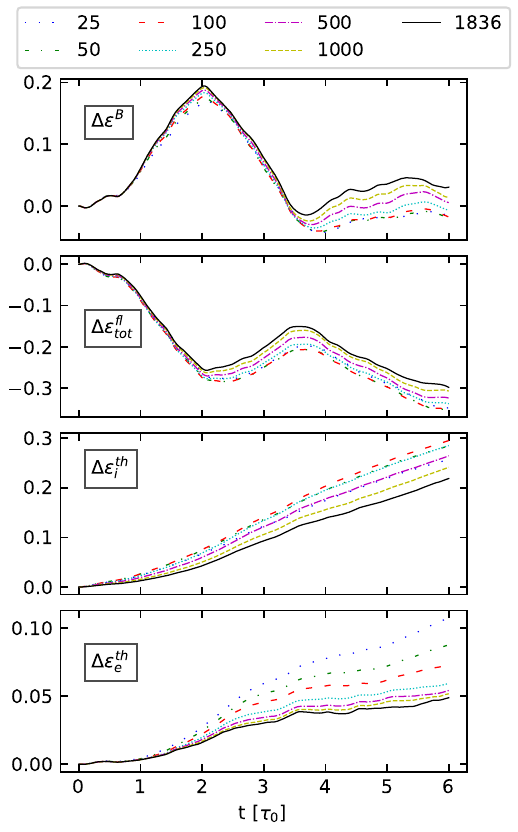}
    \caption{Time evolution of the magnetic field fluctuation energy, $\Delta \mathcal{E}^{B}$, the ion flow energy, $\Delta \mathcal{E}_{i}^{fl}$, the ion thermal energy, $\Delta \mathcal{E}_{i}^{th}$, and the electron thermal energy, $\Delta \mathcal{E}_{e}^{th}$. All values are normalised to $\mathcal{E}^{B}(0) + \mathcal{E}^{fl}_{i}(0) + \mathcal{E}^{fl}_{e}(0)$.}
    \label{fig:energies_normalised}
\end{figure}

The intense intermittent structures as well as the enhanced power at small scales affect the energetics of the plasma, resulting in very different behaviour of internal energies of ions and electrons. Fig \ref{fig:energies_normalised} shows the change in magnetic energy (top panel), flow energy (second panel), internal energy of the ions (third panel), and internal energy of the electrons (bottom panel). The energies have been normalized to the energy in magnetic and flow fluctuations at $t=0$. Alfv\'enic exchange between flow and magnetic energies typical of the Orszag-Tang vortex \citep{orszagJFM79} is visible, with the peak happening at $t\sim 2\tau_0$, consistent with existing works. Clear patterns arise with both the magnetic and the kinetic energies decaying less with increasing mass ratio. This is reflected in the reduced heating of the ions as well as the electrons. 

The heating of the ions and electrons can be divided into two regimes: i) initial phase between $ 1\tau_0 < t < 4\tau_0$, and ii) later phase $t > 4 \tau_0$. The initial phase shows rapid increase in the internal energies, fed by the intense reconnection of the central current sheet \citep[see][for examples of the central current sheet]{parasharPoP2009, grovseljAPJ17}. The system still has a large scale inhomogeneity in this phase because of the two large islands. Significant differences in the heating of ions and electrons arise in this early relatively inhomogeneous phase. At later times, the system becomes relatively more homogeneous (see Fig \ref{fig:jz} for a late time example with significantly more structure). Here, to uderstand the effects of homogeneous turbulence on heating of electrons and ions, we focus on the late times. 

\begin{figure}[h!]
    \centering
    \includegraphics[width=\columnwidth]{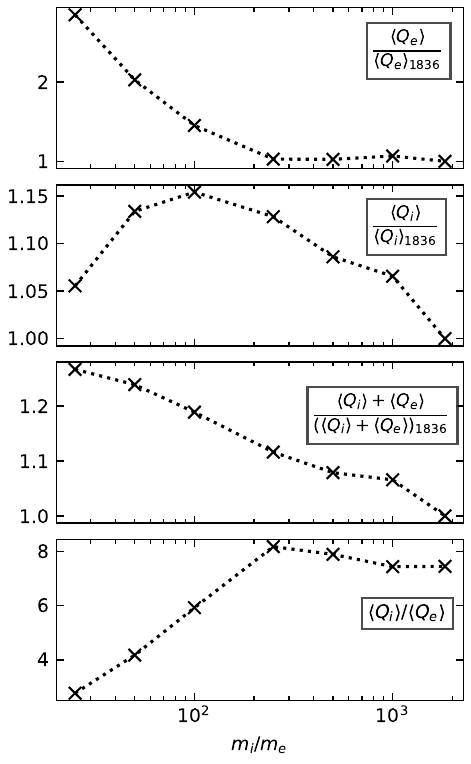}
    \caption{Showing how combinations of the mean average ion and electron heating rates change with increasing $m_i/m_e$ between 4$\tau_0$ and 6$\tau_0$. The top panel shows the average ion heating rate, $\langle Q_i \rangle$, the second to top shows the average electron heating rate, $\langle Q_e \rangle$, the third from top shows the ratio of average ion heating to average electron heating, $\langle Q_i \rangle / \langle Q_e \rangle$, and the bottom panel shows the sum of the average ion and average electron heating, $\langle Q_i \rangle + \langle Q_e \rangle$.}
    \label{fig:qi_qe_combination}
\end{figure}

We compute the heating rates ($Q = d \mathcal{E}^{th}/ dt$) for ions and electrons for the time window $4 \tau_0 < t < 6\tau_0$. Fig \ref{fig:qi_qe_combination} shows the average heating rate for electrons ($\langle Q_e\rangle$, top panel), for ions ($\langle Q_i \rangle$, second panel), total heating rate (third panel), and the relative heating rate $\langle Q_i\rangle/\langle Q_e \rangle$ for all the runs as a function of the mass ratios. All values, except for the bottom panel, have been normalized to the value for $m_i/m_e = 1836$ case to highlight how the artificial mass ratios show significantly different heating characteristics. Electron heating rate is significantly larger for smaller mass ratios, becoming as high as 300\% larger than the realistic mass ratio. This is due to the fact that the electron scales for smaller mass ratio are very close to ion scales and hence the electrons experience significantly larger fluctuations. The relatively constant power at ion scales also results in the ion heating rate changing very little with a maximum deviation of $\sim 15$\%. The total heating rate is over-predicted, due to large electron heating, by roughly 30\% in the smallest mass ratio case. Although the electron heating rate changes significantly across these simulations, the ion heating rate is always larger than the electron heating rate. However, the small mass ratio simulations under-predict the relative heating rate by more than a factor of 2. 

An interesting feature is that the electron heating rate as well as the relative heating rate approach the realistic value at $m_i/m_e = 250$. It implies that the scale separation between ions and electrons approaches a sufficiently large value at $m_i/m_e \sim 250$ to ensure that the ion-scale fluctuations do not significantly affect electron dynamics. 

\begin{figure*}[h!]
    \centering
    \includegraphics[width=7.0in]{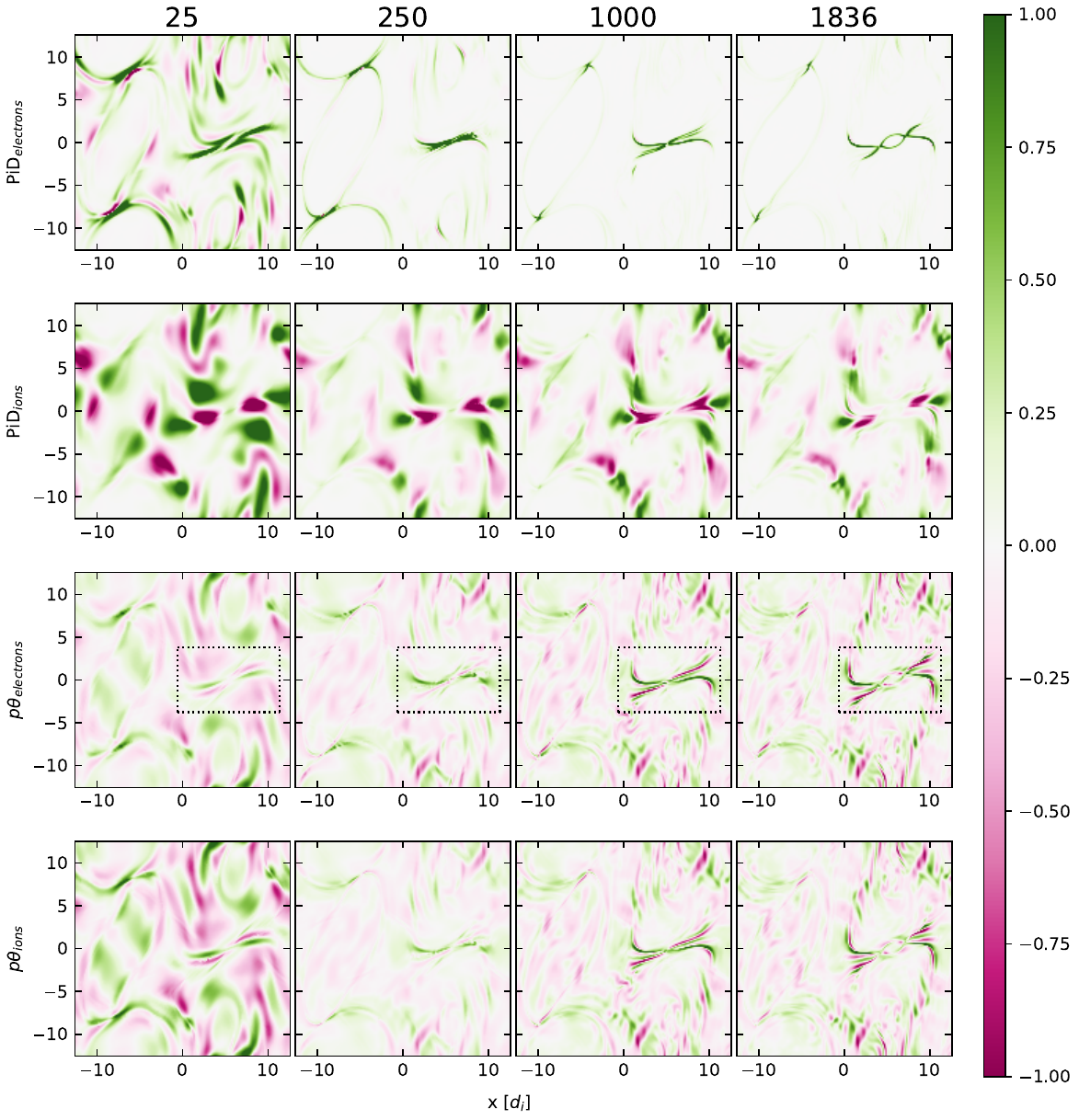}
    \caption{Showing the intensity and spatial distribution of dissipative terms during the turbulent regime, at $4.9\tau_0$. The periodic domain has been rolled by $4\pi d_i$ in the y direction and $1.76\pi d_i$ in the x direction to place the most intense current sheet near the middle for better visibility.  Top two panels; $\boldsymbol{\Pi} \boldsymbol{D}$ for electrons and ions respectively. Bottom two panels; $p\theta$ for electrons and ions respectively. The dotted regions on the $p\theta_{electrons}$ panel is shown in fig \ref{fig:pth_zoomed}}
    \label{fig:big_multi}
\end{figure*}

\begin{figure}[h!]
    \centering
    \includegraphics[width=0.8\columnwidth]{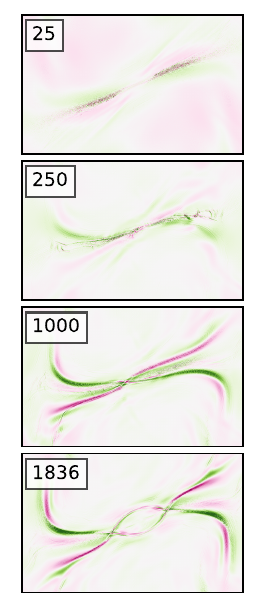}
    \caption{Showing the region of $p_e\theta_e$ enclosed by the dotted boxes in \ref{fig:big_multi}. The figure acts to qualitatively show the development of structure with increasing $m_i/m_e$, with the colours corresponding to the values shown on the colour bar in fig \ref{fig:big_multi}.}
    \label{fig:pth_zoomed}
\end{figure}

\begin{figure}[h!]
    \centering
    \includegraphics[width=\columnwidth]{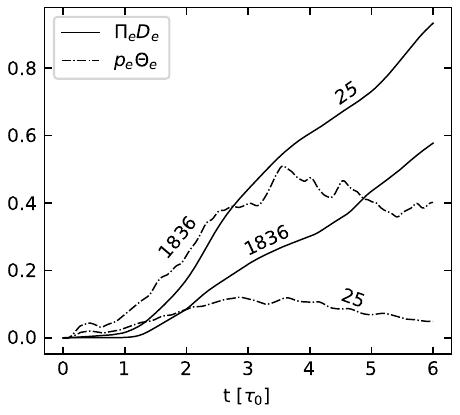}
    \caption{Showing the cumulative $\Pi_eD_e$ and $p_e\theta_e$ normalised to the final value of $\Delta\mathcal{E}_e^{th}$ for $m_i/m_e$ = 25 and 1836 .}
    \label{fig:pid_vs_pth_e}
\end{figure}

To investigate the processes affecting the changing heating rates with changing mass ratios, we analyse the pressure-strain interaction, which is responsible for the transfer of energy into the internal degrees of freedom of the plasma.
A quick manipulation of the Vlasov-Maxwell set of equations gives us (see e.g. \textcite{yangPhysRevE2017})
 
\begin{eqnarray}
    \partial_t\langle\mathcal{E}^{fl}_\alpha \rangle & = & \langle \PS \rangle + \langle \mathbf{J}_{\alpha}\cdot\mathbf{E} \rangle \nonumber \\
    \partial_{t}\langle \mathcal{E}_{\alpha}^{th} \rangle & =  & - \langle \PS \rangle \nonumber \\
    \partial_t \langle \mathcal{E}^{em}_{\alpha}\rangle & = & - \langle \mathbf{J}\cdot\mathbf{E} \rangle
\end{eqnarray}

where $\mathcal{E}^{fl}_\alpha = \frac{1}{2}\rho_{\alpha}\mathbf{u}_{\alpha}^2$ is the fluid flow energy, $\mathcal{E}_{\alpha}^{th}=\frac{1}{2}m_{\alpha}\int (\mathbf{v}-\mathbf{u}_{\alpha})^2f_{\alpha}d\mathbf{v}$ is the thermal (or random) energy, $\mathcal{E}^{em}_\alpha = \frac{1}{8\pi}(\mathbf{B}^2 + \mathbf{E}^2)$ is the electromagnetic energy, $\mathbf{J}_{\alpha} = n_{\alpha} q_{\alpha} \mathbf{u}_{\alpha}$ is the electric current density with $\mathbf{J}=\sum_{\alpha}\mathbf{J}_{\alpha}$ being the total electric current density and $\langle ... \rangle$ denoting space averages over the entire volume. The pressure strain interaction term $\PS=P_{ij}\partial_iu_j$ can be decomposed into multiple terms via the decomposition of the pressure tensor $\boldsymbol{P}_{\alpha, ij} = p_{\alpha}\delta_{ij} + \Pi_{\alpha,ij}$ as well as the decomposition of the strain tensor $\partial_i u_{\alpha, j} = \boldsymbol{S}_{\alpha,ij} + \boldsymbol{\Omega}_{\alpha,ij} = \frac{1}{3} \theta_{\alpha} \delta_{ij} + \boldsymbol{D}_{\alpha,ij} + \boldsymbol{\Omega}_{\alpha,ij}$, where $p_{\alpha}$ is the scalar pressure, $\Pi_{ij}$ is the traceless part of the pressure tensor which is symmetric, $\theta_{\alpha}$ is the divergence of the flow field (trace of the strain tensor), and $D_{\alpha,ij}$ and $\Omega_{\alpha,ij}$ are the traceless symmetric and antisymmetric parts of the strain tensor. Using these, the pressure strain interaction can be written as (keeping the negative sign to highlight the fact that it is a source term for the internal energy):

\begin{eqnarray}
    - \PS & = & - p_{\alpha}\theta_{\alpha} -  \Pi_{\alpha,ij}D_{\alpha,ij} \nonumber \\
        & = & p\theta + PiD
\end{eqnarray}

Here, the first term (nicknamed $p\theta$, with the negative sign absorbed) is compressive ``dissipation''  responsible for the heating/cooling due to compressions/rarefactions in a gas \citep{chapman1990, huang2008book}. The second term (nicknamed Pi-D, with the -negative sign absorbed) is what becomes the ``incompressible dissipation'' i.e. the viscous term in the highly collisional cases. In kinetic plasmas the second term has to be treated in its full glory to understand the change in internal energy of the plasma. Hence, these terms are also a good measure to quantify the (in)compressible nature of a given system.

Fig \ref{fig:big_multi} shows Pi-D as well as $p\theta$ for ions and electrons for the mass ratios $m_i/m_e = 25, 250, 1000, 1836$ at the late time of $t=4.9\tau_0$. The domain has been rolled in the same way as fig \ref{fig:jz} to highlight the structure around the intense current sheet. Each row has been normalized by maximum value of the variable in the $m_i/m_e=1836$ simulation divided by a number $a$, where $a=10$ for $PiD_e$, $a=2$ for $PiD_i$, $a=5.5$ for $p\theta_e$ and $a=3$ for $p\theta_i$, to highlight the structures in each row.  Two main feature stand out: i) The structures are significantly localized for electrons, as compared to ions in all simulations, and ii) the intensification and localization of the structures with increasing mass ratio is more pronounced for electrons compared to the ions.

As expected, locally the transfer of energy happens from bulk flow into internal energy (green color) and vice versa (pink color). $p\theta$ shows a lot more structure than PiD. The region of most intense activity near the current sheet shows a layered structure, showing hints of localized compressive activity. To identify the interesting patterns, we show a zoomed-in part of $\ptheta{e}$ in the same region as in fig \ref{fig:jz_zoomed}. A clear intensification of the structures and local $p\theta$ values can be seen with increasing mass ratio. For small mass ratios, the compressive heating is weak and spread over a larger region around the current sheet. As the current sheet localizes with larger mass ratios, the regions of compressive heating/cooling also start localizing. The edges of the reconnection exhaust show blobs heated by local compressions and a cooling behind them, likely because of the rarefaction caused by the tendency of the exhaust region to expand.

The domain-averaged and time-integrated PS follows the internal energy change very closely (not shown here, see \textcite{pezziPP19} and \textcite{yangApJ22} for such plots). Fig \ref{fig:pid_vs_pth_e} shows the time-integrated values of Pi-D and $p\theta$ for mass ratios 25 and 1836 for electrons, normalized to the value of $\Delta \mathcal{E}_{e}^{th}$ at $6\tau_0$ to highlight the contributions of each term to the final internal energy budget. In the $m_i/m_e=25$ case, PiD is significantly larger than $p\theta$, while the converse is true in the $m_i/m_e=1836$ case. This implies that, for electrons, incompressive viscous-like heating is dominant in the low mass ratio case, while compressive heating dominates the realistic mass ratio case. We quantify the relative importance of PiD and $p\theta$ by computing the ratio of the averages of these two quantities at late times ($4\tau_0 \le t \le 6\tau_0$) when the turbulence is fully developed. 

\begin{figure}[h!]
    \centering
    \includegraphics[width=\columnwidth]{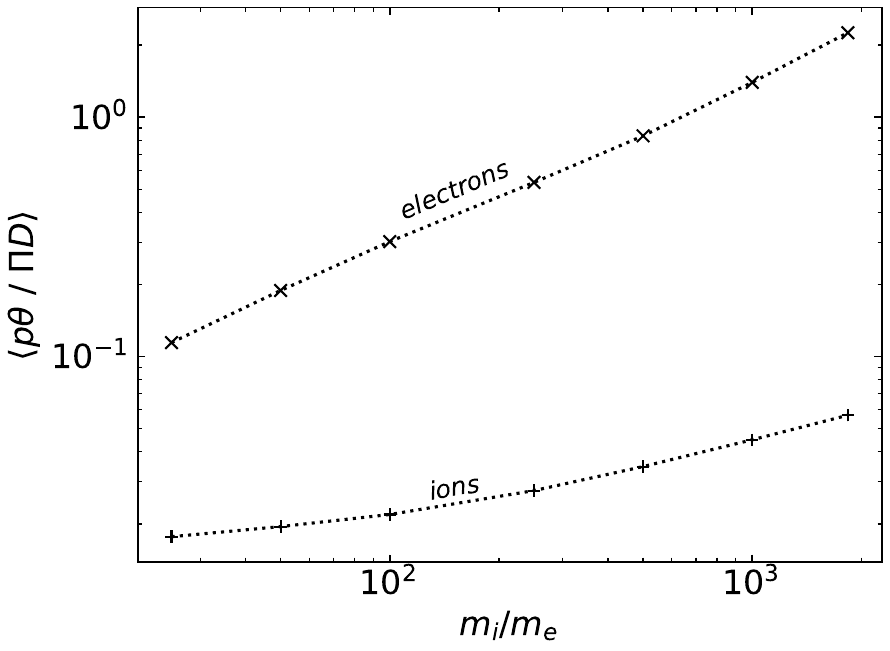}
    \caption{Shows how the compressibility changes across different simulations. To calculate the values, we take the mean average of the cumulative $p\theta$ divided by the cumulative $\Pi D$, during the turbulent regime, $4\tau_0 \leq \tau \leq 6\tau_0$.}
    \label{fig:compresibility}
\end{figure}

Fig \ref{fig:compresibility} shows the late-time average of the compressibility quantified by the ratio $p\theta$/PiD, for all simulations, for both ions and electrons. The compressibility increases with increasing mass ratio for both species. The ions stay nearly incompressible in the sense that $p\theta$ is significantly smaller than PiD in all cases. The electrons, on the other hand, go from being nearly incompressible (PiD $\sim 10p\theta$) to highly compressible ($p\theta \sim 2$ PiD). This implies that if one wants to understand the relative contributions of compressive and incompressive processes to plasma heating, a (nearly) realistic mass ratio is very desirable.

\section{Conclusion}\label{Conculsion}
Numerical simulations of kinetic plasma turbulence make compromises on some physical parameters to make the computational costs of such simulations manageable. Typical parameters that are compromised are: speed of light \citep[see e.g.][]{daughton2006PoP, parashar2016APJ}, the mass ratio of ions and electrons \citep[see e.g.][]{daughtonnature11, grovseljAPJ17}, or the system size \citep[see e.g.][]{tenbargeAPJL2013}. Such compromises have different effects on the dynamics captured. 

An artificial speed of light does not significantly affect the behaviour of linear kinetic modes \citep[see ][]{verscharenRAS2020} even up to unrealistic values of $v_A/c \sim 0.1$. If $v_A/c$ gets large enough relativistic effects need to be considered. We are not aware of any papers that have studied this in the nonlinear regime for the obvious reason of the computational cost introduced by a large speed of light. The 10 moment model might however make this study possible.  

A small system size fails to capture the inertial range dynamics and the cascade feeding into the kinetic range is either modeled or is inconsistent. This includes the failure to capture self-consistent evolution of the ion-scale structures and their role in heating of the ions \citep{parasharApJ15}. Very few studies have simulated scales from the inertial range down to the electron scales with realistic mass ratio \citep[e.g.][]{toldPRL2015}. Given the computational cost of such studies, it is nigh impossible to perform a wider parameter sweep to understand the effects of unrealistic variables with (almost) complete kinetic physics.

An unrealistic mass ratio can affect, in the linear regime, the quasi-parallel fast magnetosonic branch and the quasi-perpendicular kinetic Alfv\'en wave branch. As discussed in the introduction, some work has been done within the context of reconnection and fusion plasmas to understand the effect of unrealistic mass ratios. The artificial mass ratio affects the heat-flux at electron scales \citep{howard2015PPaCF}, and electron dynamics such as heating, energy spectra, and acceleration in magnetic reconnection \citep{li2019APJ}. The latter study however did not go up a realistic value of the mass ratio.

Within the turbulence context the paper by \cite{gary2016ApJ} is the closest candidate to our study. In their case, they fixed the system size in units of $d_e$ and changed the mass ratio. Larger mass ratios in this case resulted in ion scales ($\rho_i$ and $d_i$) that increased in an unconstrained manner, relative to the domain size. This led, we believe, to a progressive decoupling of the ions with increasing mass ratio. This resulted in the heating of the ions being affected.

In this study, we have conducted simulations of turbulence covering inertial range scales well above $d_i$ down to sub-electron scales. Special care was taken to ensure that the system size in terms of $d_i$ stayed the same, while resolving $d_e$ even for the realistic mass ratio. All simulations have the same numerical resolution, implying that the low $m_i/m_e$ cases are over-resolving the electron scales. This however, ensures a much more stringent comparison of various kinetic range dynamics, while creating the same system in the inertial range. Given the large scale-separation, the compromise made by us is in the model. Instead of using a fully kinetic model such as Vlasov-Maxwell, we use a ten-moment two-fluid model that captures the dynamics of the full pressure tensor while giving us a significant computational speed-up to allow a wider parameter space exploration.

Increasing mass ratio creates more localized and intense structures as captured by the rms current as well as the scale dependent kurtosis. This is expected because the increasing mass ratio implies smaller electron scales, giving a larger band-width for the turbulent structures to grow. The intensification of small-scale structures is reflected in the power-spectra of the magnetic field and flow velocities of both ions and electrons. The larger mass ratio cases show significantly more power at smaller scales. 

The change of structure and spectral features with mass ratio affects both the ions and electrons, especially at the early times during the Alfv\'enic exchange between flow and magnetic field. However, in the fully-developed turbulence regime at late-times, the ion-heating rate is relatively constant (with only $\sim$ 15\% error), consistent with the findings of \cite{li2019APJ} in the context of magnetic reconnection. Similar to \cite{li2019APJ}, the electron heating is significantly affected in our simulations as well. The electrons heat significantly more, with the heating rates being up to 300\% times the realistic value in the lowest mass ratio case. The heating rate approaches the realistic value at the mass ratio $m_i/m_e=250$, resulting in the relative ion/electron heating plateauing at the same mass ratio. Hence, this mass ratio is an ideal compromise if the relative heating rates are the primary concern. The dissipation channels tell a different story though.

The relative change in the heating dynamics is mediated by increasingly localized pressure-strain interactions. Not only do the quantities such as PiD and $p\theta$ become more localised, the compressive physics becomes significantly more important for the plasma heating. This effect is nominal for the ions but becomes markedly prominent for the electrons, with the electrons transitioning from being nearly incompressible at smaller mass ratios to being extremely compressible at the larger mass ratios. This effect does not show a plateauing similar to the heating rates, implying that near-realistic mass ratios are desirable if one wishes to explore the relative importance of compressive and incompressive physics in dissipation.

Much needs to be done to understand the effects of such parameter variations in the kinetic range. The most obvious being a scale-filtering analysis \citep[e.g][]{yangApJ22} to identify the scales at which ion and electron heating dominates. An analysis like this would help put in perspective disparate findings such as the heating of electrons at scales comparable to or larger than $d_i$ or the heating of ions at electron scales \citep{matthaeusApJ2020, toldPRL2015}. 

As mentioned earlier, the lack of relativistic effects, and higher order kinetic physics are two of the limitations of this study. A future study could incorporate updated closure models \citep[][e.g.]{ngPoP20} to investigate if comparisons to kinetic simulations are improved further. Another notable limitation of the study is that the simulations are ran with only two spatial dimensions not three. An extension of this problem to 3D, with the 10-moment model, while expensive, will be significantly more economical than fully kinetic models. Another important factor, albeit a bit computationally expensive, is to include relativistic effects to ensure that the nearly luminal electrons behave properly. The computational cost could be alleviated by approaching the problem in a step-by-step manor with methods for fluid-kinetic coupling, such as utilising Hermite-spectral codes \citep{delzanno2015JCP, vencels2016JoPCS, roytershteyn2018FASS,koshkarov2021CPC}.


\section{Acknowledgements}

J. Juno was supported by the U.S. Department of Energy under Contract No. DE-AC02-09CH1146 via an LDRD grant.

O.K. and G.L.D. were supported by the Laboratory Directed Research and Development Program of Los Alamos National Laboratory under project number 20220104DR. Los Alamos National Laboratory is operated by Triad National Security, LLC, for the National Nuclear Security Administration of U.S. Department of Energy (Contract No. 89233218CNA000001).

WHM and YY are partially supported for this project by NSF grant AGS-2108834 and by a NASA Living With a Star project under subcontract 655-001 from the New Mexico Consortium. 

MAS and WHM are partially supported by NASA Living With a Star project 80NSSC20K0198: LWS Coupling electron and proton kinetic physics in the solar wind. 

\bibliography{main}{}

\bibliographystyle{aasjournal}


\end{document}